\documentclass[aps,showpacs,twocolumn,prb]{revtex4-1}
\usepackage{hyperref}
\usepackage{graphicx}
\usepackage{amsmath,amssymb,amsfonts}

\begin{document}

\title{Anisotropic $H_{c2}$, thermodynamic and transport measurements, and pressure dependence of $T_{c}$ in K$_{2}$Cr$_{3}$As$_{3}$ single crystals}
\author{Tai Kong}  
\author{Sergey L. Bud'ko}  
\author{Paul C. Canfield}
\affiliation{Ames Laboratory, U.S. DOE, Iowa and Department of Physics and Astronomy, Iowa State University, Ames, Iowa 50011, USA}

\begin{abstract}
We present a detailed study of single crystalline K$_{2}$Cr$_{3}$As$_{3}$ and analyze its thermodynamic and transport properties, anisotropic $H_{c2}(T)$, and initial pressure-dependence of $T_{c}$. In zero-field, the temperature-dependent resistivity is metallic. Deviation from a linear temperature-dependence is evident below 100 K and a $T^{3}$-dependence is roughly followed from just above $T_{c}$ ($\sim$10 K) to $\sim$ 40 K. Anisotropic $H_{c2}(T)$ data were measured up to 140 kOe with field applied along and perpendicular to the rod-like crystals. For the applied field perpendicular to the rod, $H_{c2}(T)$ is linear with a slope $\sim$-70 kOe/K. For field applied along the rod, the slope is about -120 kOe/K below 70 kOe. Above 70 kOe, the magnitude of the slope decreases to $\sim$-70 kOe/K. The electronic specific heat coefficient, $\gamma$, just above $T_{c}$, is 73 mJ/mol K$^{2}$; the Debye temperature, $\Theta_{D}$, is 220 K. The specific heat jump at the superconducting transition $\Delta$C $\sim$ 2.2 $\gamma T_{c}$. Finally, for hydrostatic pressures up to $\sim$7 kbar, $T_{c}$ decreases under pressure linearly at a rate of -0.034 K/kbar. 
\linebreak
\linebreak
\end{abstract}
\maketitle

K$_{2}$Cr$_{3}$As$_{3}$ has been recently discovered as a new superconducting material with a $T_{c}$ of 6.1 K\cite{Bao14}. Having a structure that contains (Cr$_{3}$As$_{3}$)$^{2-}$ chains, it quickly arouses interest as a potentially new, quasi-one-dimensional (Q1D) superconductor. Related compounds, Rb$_{2}$Cr$_{3}$As$_{3}$ and Cs$_{2}$Cr$_{3}$As$_{3}$, with $T_{c}$ values of about 4.8 K and 2.2 K respectively were reported soon after the discovery of K$_{2}$Cr$_{3}$As$_{3}$\cite{Tang14, Tang15}. Band structure calculation was also conducted and compared with experimental results\cite{Jiang14}. In the initial report\cite{Bao14}, data were acquired primarily on polycrystalline samples of K$_{2}$Cr$_{3}$As$_{3}$. For Q1D superconductors, it is believed that with magnetic field applied along the chain, the upper critical field, $H_{c2}$, should be much higher than that when the field is along other directions due to minimized orbital pair breaking. Experimentally, this was observed for Q1D systems like Li$_{0.9}$Mo$_{6}$O$_{17}$ and (TMTSF)ClO$_{4}$\cite{Mercure12,Yonezawa08}. In this paper, we present detailed anisotropic $H_{c2}(T)$ data via resistance measurements on K$_{2}$Cr$_{3}$As$_{3}$ single crystals. In addition, with the recent discovery of iso-structural compound, Rb$_{2}$Cr$_{3}$As$_{3}$, with a larger unit cell and a lower $T_{c}$ value, it is interesting to study the pressure-dependence of $T_{c}$ of K$_{2}$Cr$_{3}$As$_{3}$. Results from low-field dc magnetization measurements with pressures up to 7 kbar will be presented.

\begin{figure}[tbh!]
\includegraphics[scale = 0.32]{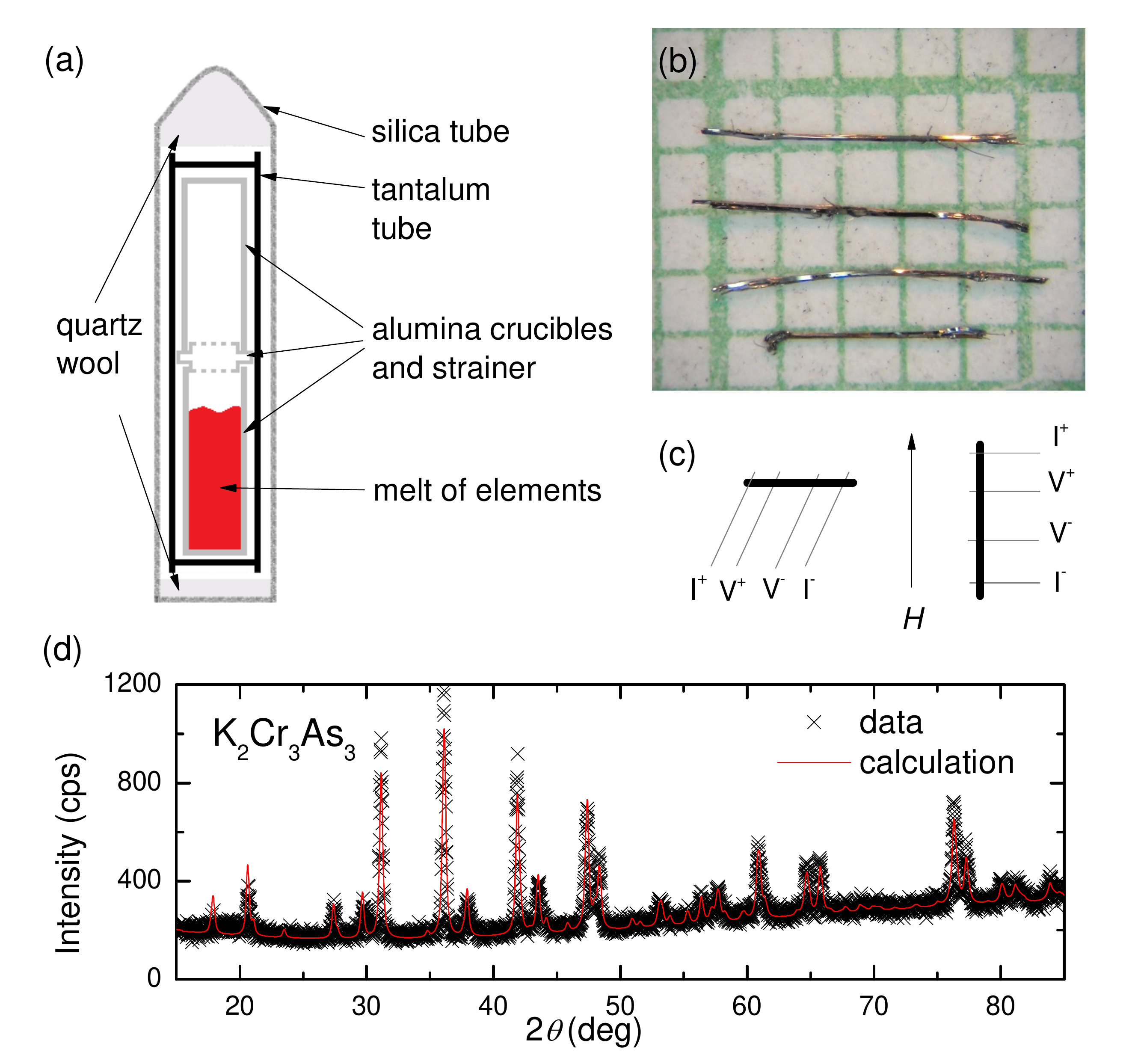}
\caption{(Color online) (a) A schematic drawing of the ampoule assembly for the material synthesis. (b) Typical habit of the single crystals of K$_{2}$Cr$_{3}$As$_{3}$ on a millimeter grid paper. (c) The contacts arrangements and external field direction for the resistance measurements. (d) X-ray diffraction data on ground/deformed crystals. The black crosses represent the experimental data and the red solid line represents the calculation based on the crystallographic information. For details of the crystal structure, readers are directed to Ref.~\onlinecite{Bao14}.}
\label{1}
\end{figure}

K$_{2}$Cr$_{3}$As$_{3}$ single crystals were grown using a high-temperature solution growth method\cite{Canfield92}. A schematic drawing of the Matryoshka-like ampoule assembly for crystal synthesis is shown in Fig.~\ref{1}(a). Elemental K, Cr, and As (in bulk/lump form) were packed in an alumina crucible following the ratio listed in Ref.~\onlinecite{Bao14} (K:Cr:As = 6:1:7). Crucibles and starting material were then welded into a tantalum tube and sealed in a silica ampoule under partial argon atmosphere. To avoid possible explosion due to the high vapor pressure of the un-reacted elements, the whole ampoule was slowly heated up to 1000 $^{\circ}$C over 2 days. It was then cooled down over $\sim$ 100 hrs to 700 $^{\circ}$C, at which temperature the single crystals and the remaining liquid were separated in a centrifuge by the alumina strainer that was placed in between the growth side crucible and an empty (catch) crucible. The crystals obtained are rod-like (Fig.~\ref{1}(b)) and malleable. A rough powder x-ray diffraction pattern was collected on ground/deformed crystals using a Rigaku Miniflex unit located in a N$_{2}$ glove box. The data, along with a fit to the crystallographic data given in Ref.~\onlinecite{Bao14} are shown in Fig.~\ref{1}(d) and are consistent with the single crystals adopting the hexagonal, a = 9.983 $\AA$, c = 4.230 $\AA$ unit cell. Given the unit cell dimensions and single crystal morphology, we identify the rod direction as being along the crystallographic c-axis. 

Resistance was measured using a standard 4-probe technique. DuPont 4929N silver paint was used to attach platinum wires onto the sample in a N$_{2}$ glove box. Electric current was applied along the rod for anisotropic $H_{c2}$ measurements (Fig.~\ref{1}(c)). Long, straight samples were used and supported by flat plastic pads for all resistivity measurements to avoid potential torque that could deform the crystals and thus change their alignment. Resistivity was estimated assuming the sample has a cylindrical shape. The absolute value of the resistivity is therefore only accurate to within a factor of three. The temperature and field-dependent resistance was measured in a Quantum Design (QD) Physical Property Measurement System, PPMS-14 ($T$ = 1.8-305 K, $H$ = 0-140 kOe). Specific heat data were measured using a QD PPMS via relaxation method. A $^{3}$He option was utilized to obtain specific heat data down to 0.4 K. Specific heat data were measured on a "raft" assembled out of several single crystals. Despite the fact that samples' shape and arrangement was not ideal for such measurements, the coupling constant was high, between 96$\%$ and 100$\%$.

The pressure-dependence of the superconducting transition temperature, $T_{c}$, was determined via low-field (20 Oe), zero-field-cooled (ZFC), dc magnetization measurements in a QD Magnetic Property Measurement System (MPMS) using a commercial, HMD, Be-Cu piston cylinder pressure cell\cite{MPMScell}. Daphne oil 7373 was used as the pressure medium and Pb was used as the manometer\cite{Schilling81}. Since the samples are very air sensitive, exposure to air was limited as much as possible to avoid oxidation.

\begin{figure}[tbh!]
\includegraphics[scale = 0.32]{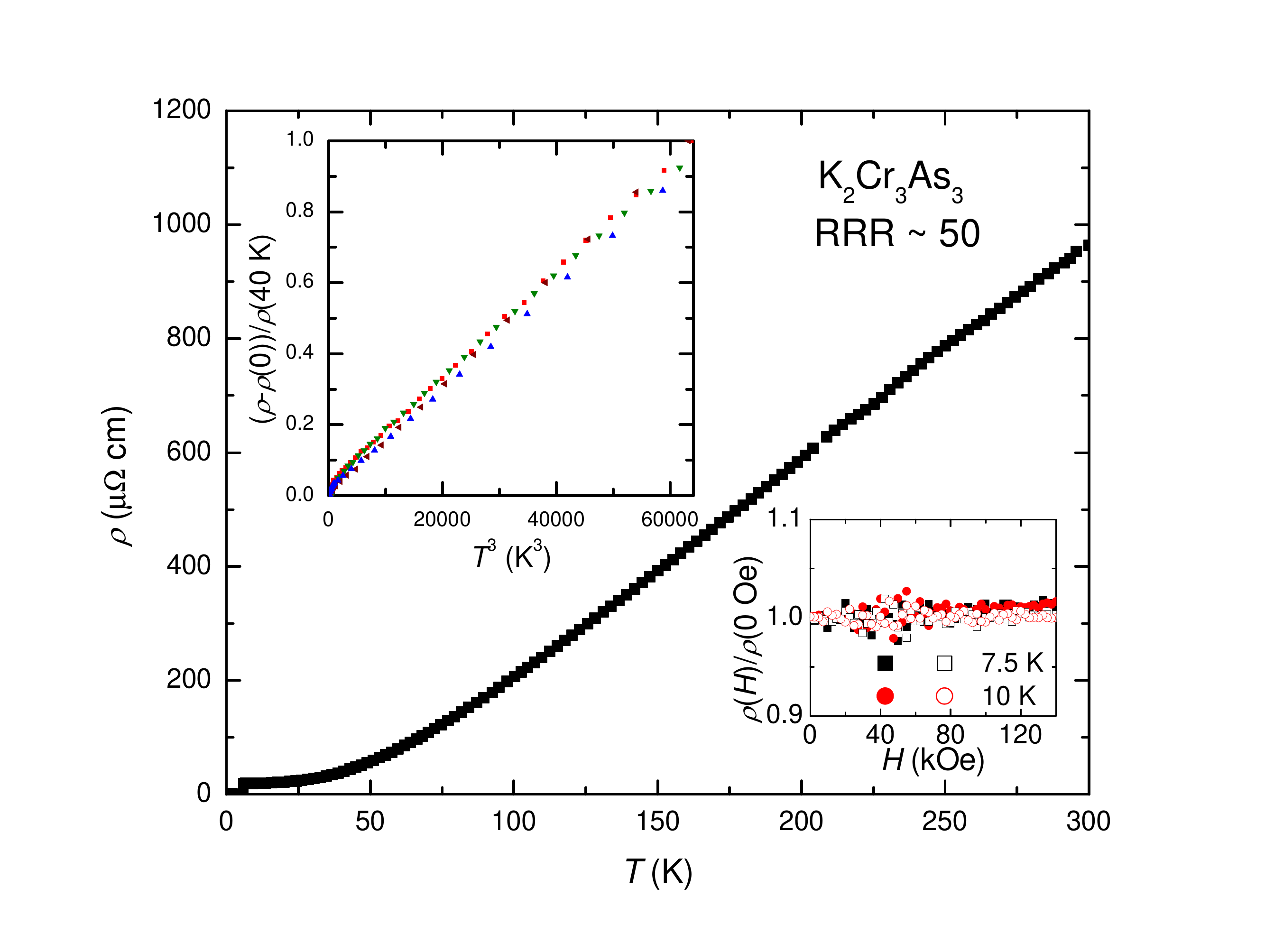}
\caption{(Color online) Temperature-dependent resistivity of K$_{2}$Cr$_{3}$As$_{3}$. The left inset shows the normalized resistances versus $T^{3}$ below 40 K for 4 different samples. The right inset shows the magnetoresistance measured at 7.5 K and 10 K. Solid and hollow symbols represent transverse and longitudinal magnetoresistance respectively.}
\label{2}
\end{figure}

The temperature-dependence of the resistivity of K$_{2}$Cr$_{3}$As$_{3}$ is metallic as shown in Fig.~\ref{2}. The residual resistance ratio (RRR) is about 50, a factor of 5 better tan the RRR$\sim$10 found for the polycrystalline data\cite{Bao14}. A clear, sharp, transition at 6.1 K indicates the superconducting transition. At room-temperature, the resistivity is of the same order of magnitude as the reported value\cite{Bao14}. However, in contrast to the reported linear temperature-dependence from 7 to 300 K in Ref.~\onlinecite{Bao14}, we see a clear deviation from linearity below $\sim$ 100 K. For the four samples we have measured, we get slopes of 3.0, 2.8, 3.1 and 3.1 from a log($\rho$-$\rho_{0}$) vs. log$T$ plot between 10 and 40 K. Therefore in the left inset of Fig.~\ref{2}, we plot the resistance data from several samples normalized to their value at 40 K as a function of $T^{3}$. It is clear that a $T^{3}$ power law describes these data well over the temperature region that is shown. Slight deviation below 10 K might due to the proximity to the superconducting transition. Although we did not see a temperature region where Fermi liquid behavior is dominant, this could due to the fact that the measurement temperature is not low enough comparing with its Debye temperature. As will be discussed below, due to the extremely large $H_{c2}$ values associated with this compound, measuring normal state resistivity to low temperatures will require extremely high magnetic fields.

In the normal state, despite a relatively large RRR value, virtually no magnetoresistance was observed. The right inset of Fig.~\ref{2} shows the normalized resistance as a function of applied field for both transverse ($H \perp I$) and longitudinal ($H \parallel I$) directions of applied field. Right above the $T_{c}$, the resistance stays close to constant up to 140 kOe. Given this null response, an upper limit of 2$\%$ (at 140 kOe) can be set on the low-temperature magnetoresistance.

\begin{figure}[tbh!]
\includegraphics[scale = 0.32]{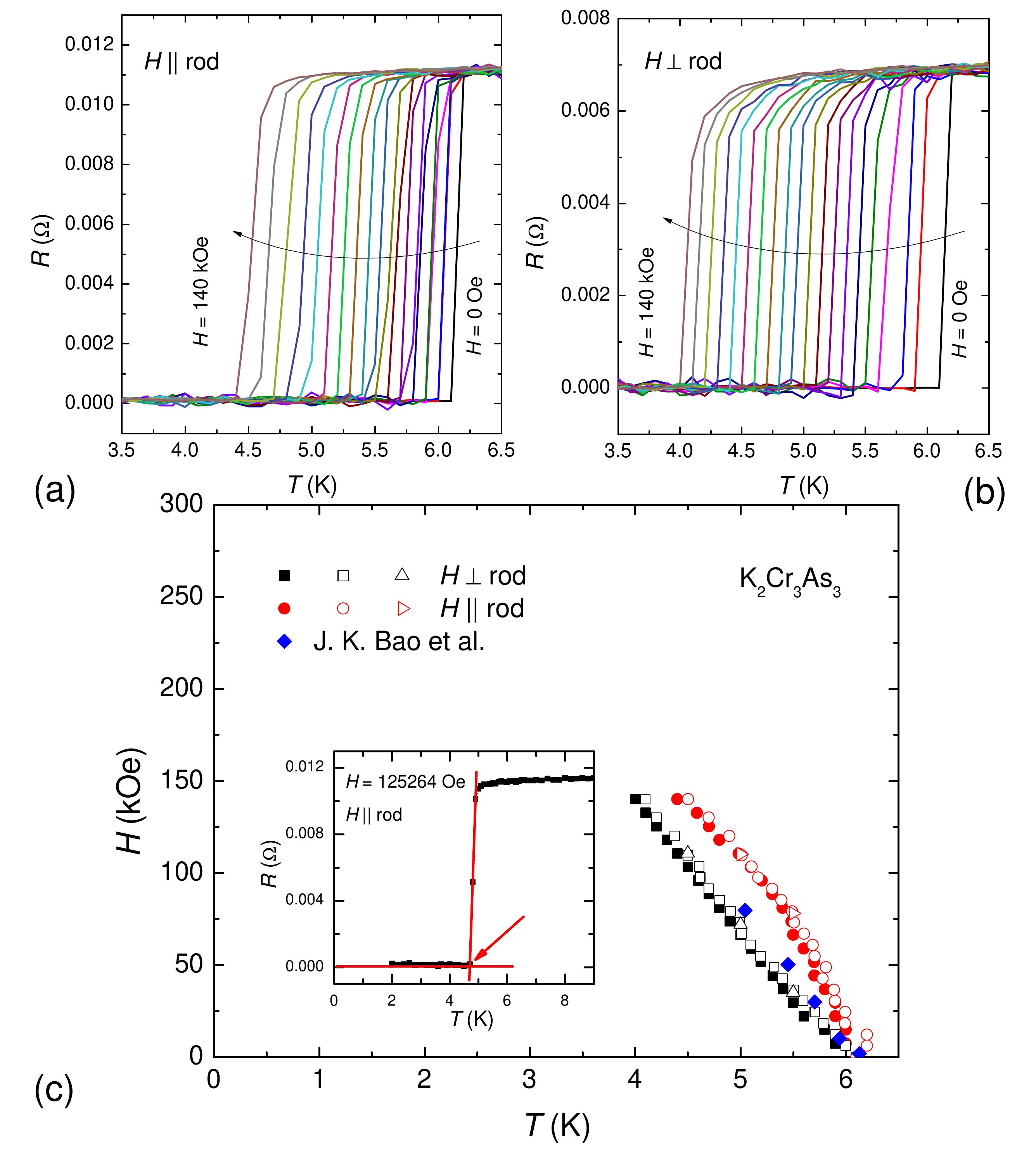}
\caption{(Color online) (a)/(b) $R$($T$) data measured for $H \parallel$ rod/$H \perp$ rod from 0 Oe up to 140 kOe in 20 uniform field steps ($\Delta H$ = 7368.4 Oe). (c) Anisotropic $H_{c2}$ of K$_{2}$Cr$_{3}$As$_{3}$. Black and red symbols represent data for $H \parallel$ rod and $H \perp$ rod. Blue points are from Ref.~\onlinecite{Bao14}. Solid and open symbols show data from two different batches of samples. Triangles are data from field-sweep at fixed temperatures. Squares and circles are data from temperature sweeps at fixed field. The inset shows a typical superconducting transition measured at 125264 Oe with red solid lines and arrow indicating the criteria for determining the transition temperature (see more in text).}
\label{3}
\end{figure}

Fig.~\ref{3}(a) and \ref{3}(b) show the temperature-dependent resistance, $R$($T$), measured under different magnetic field for $H \parallel$ rod and $H \perp$ rod respectively. Fig.~\ref{3}(c) shows the anisotropic $H_{c2}(T)$ data for K$_{2}$Cr$_{3}$As$_{3}$ from several samples. The superconducting transition is very sharp in our resistivity measurements. The inset to Fig.~\ref{3}(c) shows that, at over 125 kOe, the onset and offset of the transition differ by less than 0.25 K as opposed to an over 0.6 K width at 80 kOe for polycrystalline data\cite{Bao14}. An offset criterion\cite{Ni08Ba} was used to determine $T_{c}$ (see the inset of Fig.~\ref{3}(c)). For field applied perpendicular to the rod, the $T_{c}$ decreases almost linearly with increasing magnetic field. The slope is $\sim$ -70 kOe/K. For field applied along the rod, the initial slope is roughly -120 kOe/K, and above 70 kOe the slope is close to -70 kOe/K, which is similar to that when field is applied perpendicular to the rod. Different batches of samples as well as field-sweep data at fixed temperature give consistent results. In Fig.~\ref{3}(c), data from Ref.~\onlinecite{Bao14} are plotted as blue solid diamonds. It is close to what we obtained for field perpendicular to the rod and the slope is consistent with our current results. 

From one-band BCS theory, for a s-wave, isotropic material, in the clean limit, the orbital $H^{orb}_{c2}$ = -0.73$\mid$d$H_{c2}/dT\mid_{T_{c}}T_{c}$\cite{Tinkham}. In our case, for field perpendicular to the rod, $H^{orb}_{c2} \sim$ 312 kOe. The Pauli limit for a simple BCS superconductor is $H^{p}$ = 1.84$T_{c}$ = 110 kOe. The $H_{c2}$ of real systems will be influenced by both $H^{orb}_{c2}$ and $H^{p}$. The Maki parameter\cite{Maki66} that describes the relative importance of the two critical field limits, $\alpha = \sqrt{2} H^{orb}_{c2}/H^{p}$, is equal to 4. This suggests that at low-temperature, the $H_{c2}$ might be Pauli limited. With the current data up to 140 kOe that exceeds the one-band BCS estimated Pauli limit, we see no clear sign of saturation of $H_{c2}$. Note that, with multi-band as well as varying coupling strength, the Pauli limit can be different from one-band BCS prediction. For example, in many Fe-based superconductors, $H^{p}$ values are significantly enhanced\cite{Khim11}.

\begin{figure}[tbh!]
\includegraphics[scale = 0.32]{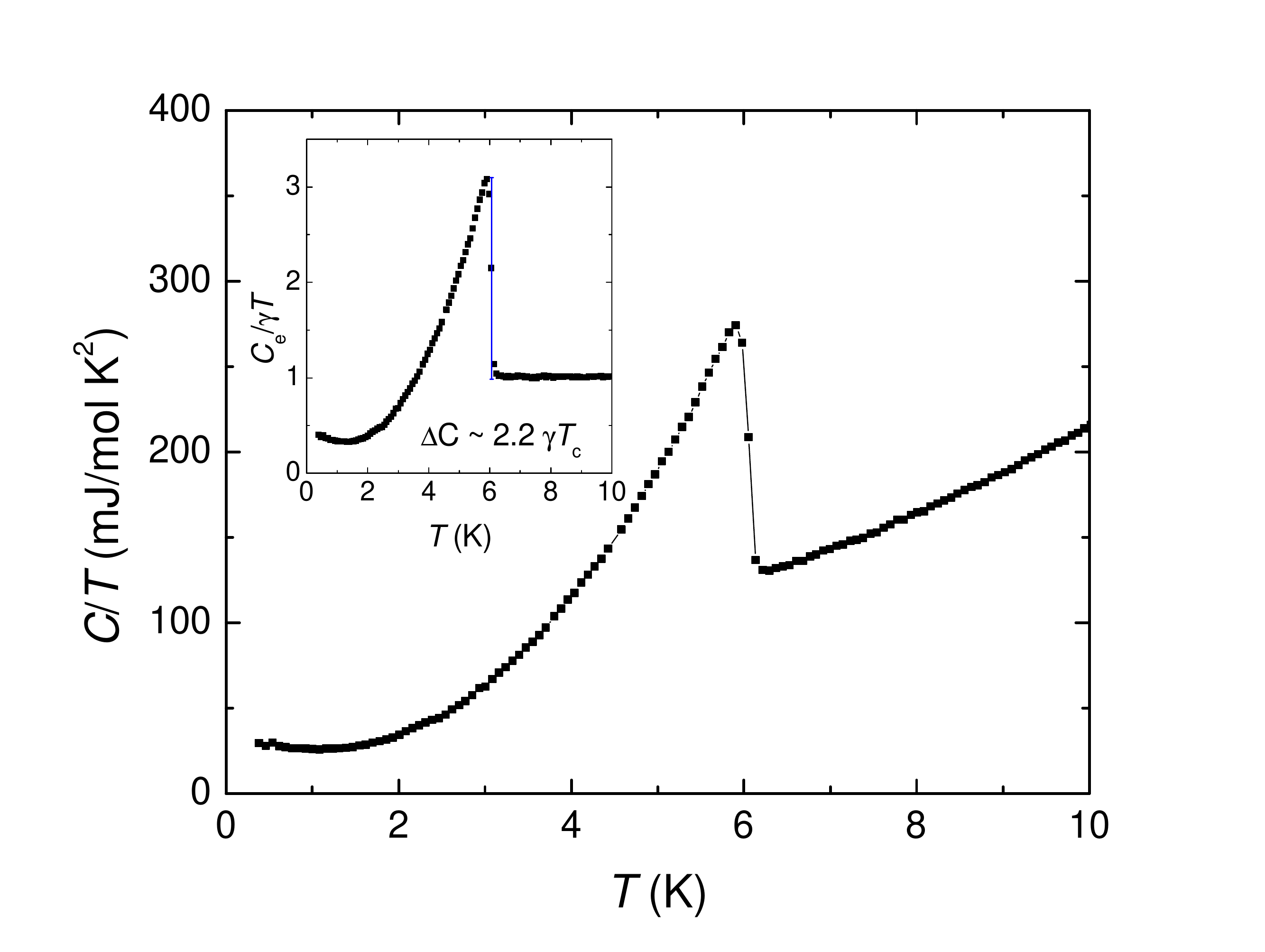}
\caption{(Color online) $C$/$T$ plotted as a function of $T$ down to 0.4 K. Inset shows the $C_{e}$/$\gamma T$ as a function of $T$. The electronic part of the specific heat, $C_{e}$, was obtained by subtracting $\beta T^{3}$ from the total specific heat.}
\label{4}
\end{figure}

The specific heat data from K$_{2}$Cr$_{3}$As$_{3}$ are shown in Fig.~\ref{4}. A clear jump in the specific heat at around 6 K corresponds to the superconducting transition. In the normal state, $C = \gamma T + \beta T^{3}$ fits the data quite well. In the range of 7-10 K, we obtained a $\gamma$ $\sim$ 73 mJ/mol K$^{2}$ and the Debye temperature, $\Theta_{D}$, derived from $\beta$, is around 220 K. These data are consistent with recently reported values\cite{Bao14}. At the superconducting transition, the specific heat jump is roughly 2.2 $\gamma T_{c}$. This is larger than the simple s-wave BCS prediction and comparable to the value obtained in Ref.~\onlinecite{Bao14}. Possibly, strong coupling is involved\cite{Carbotte90}. Below 1.5 K, we observe a clear upturn in $C/T$. Assuming that the normal state values of $\gamma$ and $\beta$ stay constant below $T_{c}$, this upturn exacerbates the difficulty of having entropy conserved, which was already hinted at in the previously reported data down to 2 K\cite{Bao14}. Additional, non-superconducting contributions, such as impurities or nuclear Schottky anomaly, may be responsible for this low-temperature rise. But whatever its origin is, it makes fitting of the data to specific models problematic.

\begin{figure}[tbh!]
\includegraphics[scale = 0.32]{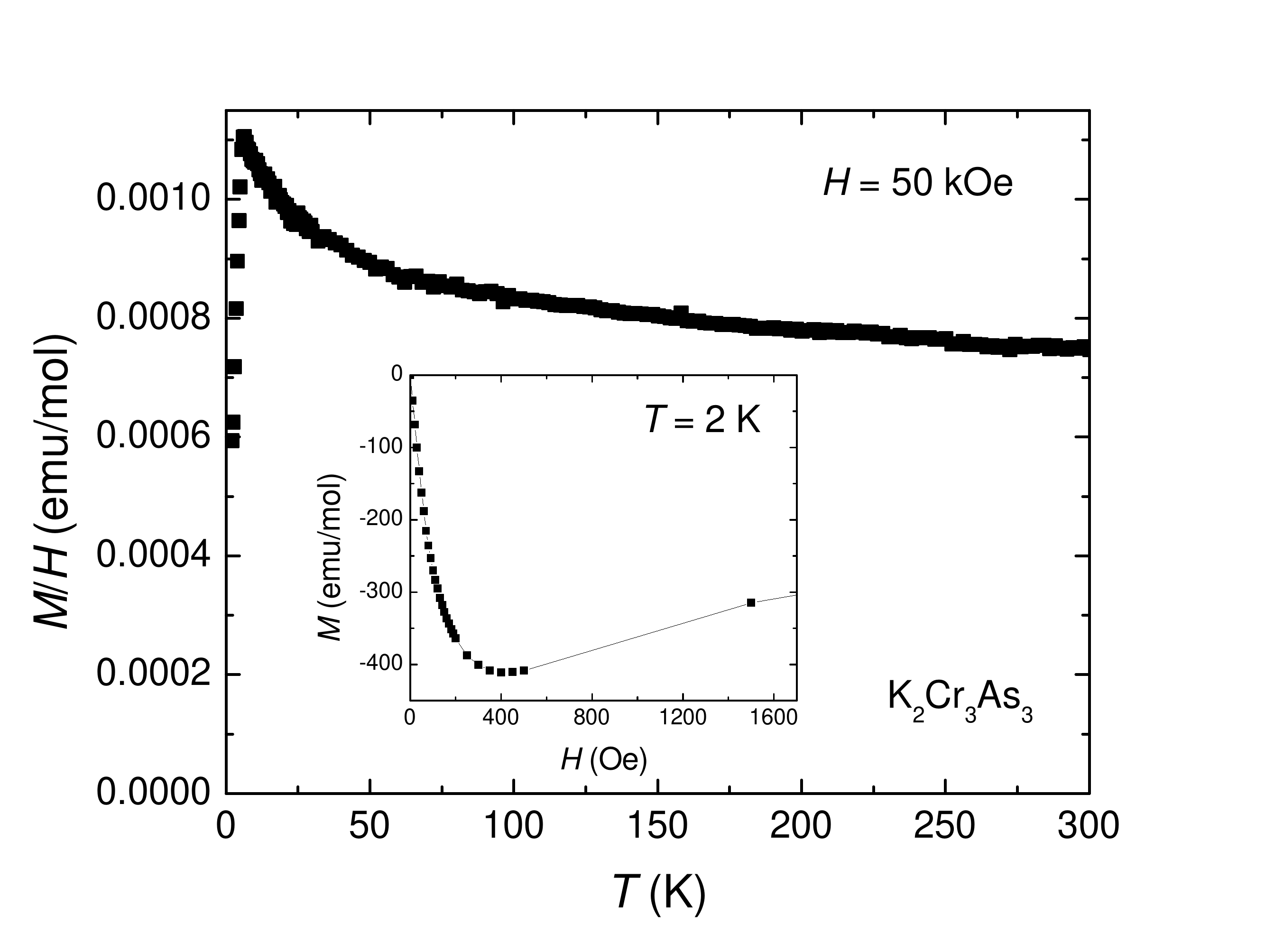}
\caption{The magnetic susceptibility measured at 50 kOe on a randomly oriented collection of single crystals. Inset shows the magnetization isotherm measured at 2 K up to 1500 Oe.}
\label{5}
\end{figure}

At ambient pressure, the normal state dc magnetic susceptibility ($\chi$ = $M$/$H$) is $\sim$ 0.9$\times$10$^{-3}$ emu/mol and decreases by more than 30$\%$ from 7 K to 300 K (Fig.~\ref{5}). If we fit these data to a Currie-Weiss law, we find a temperature-independent susceptibility $\chi_{0} \sim$ 0.7$\times$10$^{-3}$ emu/mol, a small effective moment of 0.36 $\mu_{B}$/f.u. (or 0.21 $\mu_{B}$/Cr) and a Curie-Weiss temperature, $\Theta \sim$ -40 K. At this point in time, it is not clear if such a local-moment fit is appropriate or intrinsic for this material. Given that $\gamma$ = 9 mJ/mol-atom K$^{2}$ is relatively high, it is appropriate to contrast it with the magnetic susceptibility. Assuming spin = 1/2, the estimated Wilson ratio\cite{Stewart84}, $R_{W}$, is in a range of 0.9-1.1 by taking the low-temperature magnetic susceptibility value. If we consider the Curie tail as a result of impurity contribution and thus take the value of temperature-independent $\chi_{0}$, $R_{W}$ is about 0.7. These values are close to what one would expect for a Fermi-liquid system ($R_{W} = 1$). 

\begin{figure}[tbh!]
\includegraphics[scale = 0.32]{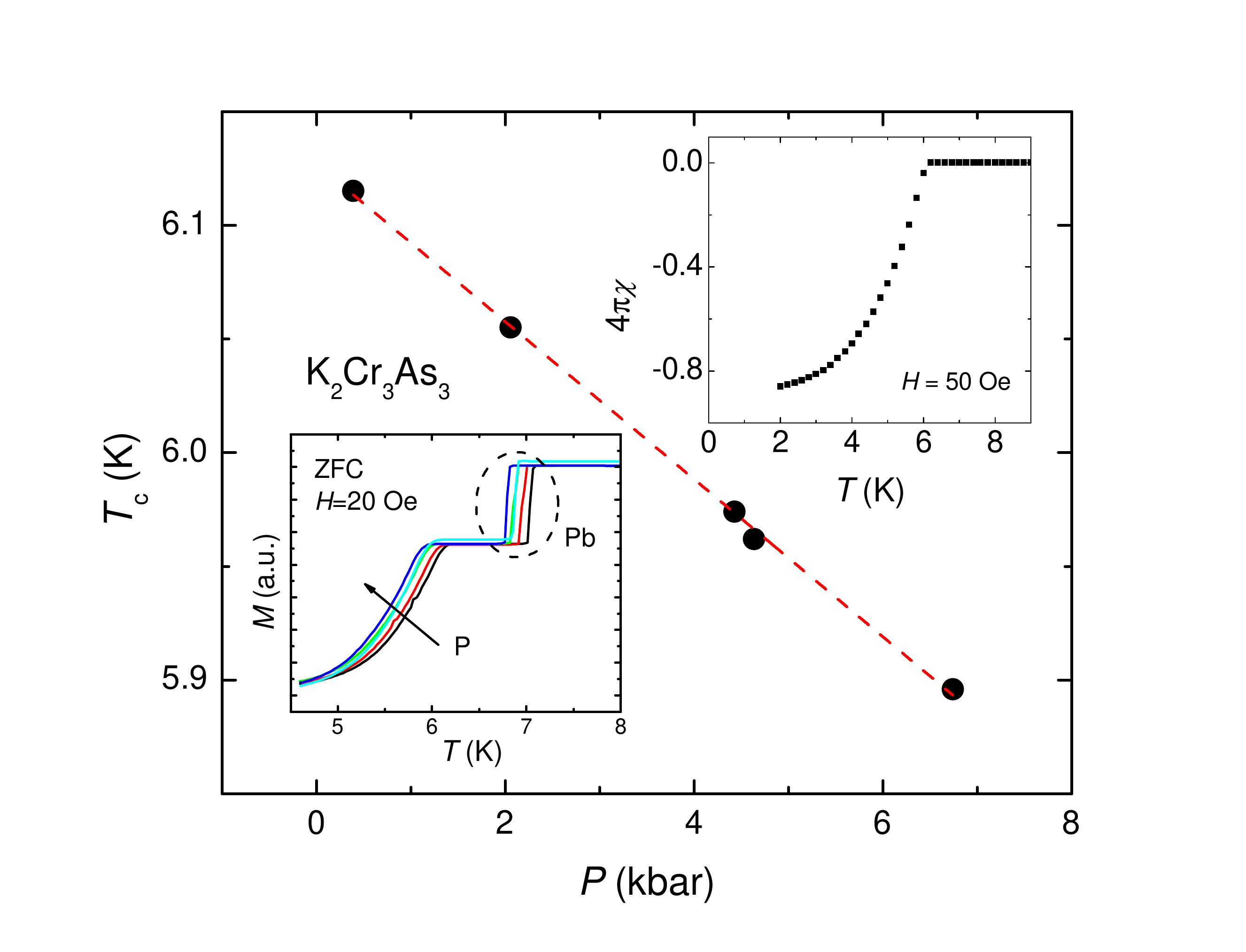}
\caption{(Color online) Pressure-dependence of $T_{c}$. Dotted line is a guide for the eye. The left inset shows the raw data with Pb serving as a manometer. The right inset shows the zero-field-cooled magnetization data under ambient pressure measured at 50 Oe.}
\label{6}
\end{figure}

Zero-field-cooled dc magnetization data are shown in the right inset of Fig.~\ref{6}. The superconducting transition appears sharply at 6.1 K, consistent with both resistance and specific heat measurements. To avoid oxidation, which could affect both sample mass and superconducting volume fraction, the sample mass was measured in a glove box and exposure to air was minimized.

At 2 K, the low-field $M$($H$) data deviate from linear field-dependence at around 70 Oe (Fig.~\ref{5}, inset). The minimum in magnetization appears at around 400 Oe. Taking $H_{c1}$ = 70 Oe and $H_{c2} \sim H^{orb}_{c2}$ = 312 kOe and using $H_{c1}$/$H_{c2}$ = ln($\kappa$)/2$\kappa^{2}$, the GL parameter, $\kappa$, is $\sim$100. From the specific heat jump at $T_{c}$ using the Rutgers relation: $\Delta$C/$T_{c}$ = (1/8$\pi \kappa^{2}$)(d$H_{c2}/dT$)$^{2}\mid_{T_{c}}$\cite{Welp89}, we obtain a similar $\kappa$ value of 116. This high value of $\kappa$ suggests that K$_{2}$Cr$_{3}$As$_{3}$ is deep in the type II regime.

The pressure-dependence of $T_{c}$ is plotted in Fig.~\ref{6}. The superconducting transition of Pb was used to determine the pressure within the pressure cell\cite{Schilling81}. As shown in the left inset of Fig.~\ref{6}, both superconducting transitions from the Pb and the sample are clear. Up to 7 kbar, the $T_{c}$ value for K$_{2}$Cr$_{3}$As$_{3}$ decreases linearly with increasing pressure at a rate of -0.034 K/kbar. Given that the larger-unit-celled Rb$_{2}$Cr$_{3}$As$_{3}$ undergoes a superconducting transition at a slightly lower temperature, 4.8 K, the effects of chemical and physical pressure are clearly not equivalent in this material.

In conclusion, we grew single crystals of K$_{2}$Cr$_{3}$As$_{3}$ using a high-temperature solution growth method. Being able to decant the molten solution at high-temperature allows us to get relatively large, free standing, high-quality single crystals, which also enables us to perform all of our measurements on single crystals or their arrays. In contrast to previously reported, linear temperature dependent resistivity from 300 K down to 7 K, resistivity of single crystal, with a RRR of 50 (RRR$\sim$10 for polycrystalline samples\cite{Bao14}), deviates from linear temperature-dependence below 100 K and roughly follows $T^{3}$ over the range of 10-40 K. Whereas the differences between the single crystalline and polycrystalline sample transport are easily explained by grain-boundary scattering and potential resistive anisotropies, the rather clear $T^{3}$ power law that we observe will require more experimental and theoretical investigation. Despite the fact that the temperature-dependence is different, if the RRR values reflect the intrinsic quality of the samples, and are not due to grain boundary scattering, the independence of $T_{c}$ on the value of RRR, or impurity scattering, suggests a conventional mechanism of its superconductivity. Anisotropic $H_{c2}(T)$ data up to 140 kOe were obtained from resistivity measurements. Up to 140 kOe, for field applied perpendicular to the rod, $H_{c2}(T)$ is linear with a slope of -70 kOe/K. For field along the rod, d$H_{c2}/dT\mid_{T_{c}}$ is about -120 kOe/K. Above 70 kOe, the the slope decreases to around -70 kOe/K. The $T_{c}$ as well as the slope of $H_{c2}$ is close to previously reported data\cite{Bao14}. The anisotropy in $H_{c2}$ is not as large as one would generally expect for a Q1D superconductor. Virtually zero magnetoresistance was observed at temperatures right above the $T_{c}$. The electronic specific heat, $\gamma$, is 73 mJ/mol K$^{2}$ or 9 mJ/mol-atom K$^{2}$. Although this is a relatively large value, the equally enhanced magnetic susceptibility leads to a Wilson ratio that is close to 1 and suggests Fermi-liquid-like properties. At ambient pressure, $H_{c1}$ is close to 70 Oe, which result in a GL parameter that is $\sim$100, taking 312 kOe as the estimated $H_{c2}$. For pressures up to 7 kbar, the superconducting transition temperature, $T_{c}$, decreases linearly with a rate of -0.034 K/kbar. Comparing to the newly discovered Rb$_{2}$Cr$_{3}$As$_{3}$, it seems to suggest that physical pressure and chemical pressure have different effect on the superconductivity in these compounds. 

Despite the fact that K$_{2}$Cr$_{3}$As$_{3}$ can be considered to be close to Q1D in a crystallographic sense, the "small" anisotropy in $H_{c2}$ appears to suggest a rather 3D nature. In addition, the exotic pairing symmetry requires more detailed study.

\section*{Acknowledgements}
We would like to thank U. S. Kaluarachchi, A. E. B$\ddot{\text{o}}$hmer, D. K. Finnemore, V. G. Kogan and V. Taufour for useful discussions. This work was supported by the U.S. Department of Energy (DOE), Office of Science, Basic Energy Sciences, Materials Science and Engineering Division. The research was performed at the Ames Laboratory, which is operated for the U.S. DOE by Iowa State University under contract NO. DE-AC02-07CH11358.

\bibliographystyle{apsrev4-1}
%

\end{document}